# A Sinogram Inpainting Method based on Generative Adversarial Network for Limited-angle Computed Tomography

**Ziheng Li,**[a] **Wenkun Zhang,**[a] **Linyuan Wang,**[a] **Ailong Cai,**[a] **Ningning Liang,**[a] **Bin Yan,**[a] **Lei Li,**[a,*]

[a]National Digital Switching System Engineering & Technological Research Centre, Zhengzhou, Henan, China, 450002

**Abstract**. Limited-angle computed tomography (CT) image reconstruction is a challenging reconstruction problem in the fields of CT. With the development of deep learning, the generative adversarial network (GAN) perform well in image restoration by approximating the distribution of training sample data. In this paper, we proposed an effective GAN-based inpainting method to restore the missing sinogram data for limited-angle scanning. To estimate the missing data, we design the generator and discriminator of the patch-GAN and train the network to learn the data distribution of the sinogram. We obtain the reconstructed image from the restored sinogram by filtered back projection and simultaneous algebraic reconstruction technique with total variation. Experimental results show that serious artifacts caused by missing projection data can be reduced by the proposed method, and it is hopeful to solve the reconstruction problem of 60° limited scanning angle.

*****Corresponding Author**, E-mail: leehotline@aliyun.com

## 1 Introduction

Limited-angle tomography is a common problem in many applications from the digital breast tomosynthesis to the industrial non-destructive testing[1]. Image reconstruction from limited-angle projections can be treated as an ill-posed inverse problem which is because of the available sinogram angle range being less than 180°. So, image reconstruction from a limited-angle scanning is challenging in computed tomography (CT) imaging. In the past decades, various methods for limited-angle CT image reconstruction can be roughly divided into two categories: image domain reconstruction methods and sinogram domain reconstruction methods.

When the missing scan angle data exceeds half of the full-angle projection data, it is difficult to obtain good reconstruction results by above traditional methods. The data-driven deep learning methods have a lot of data as the prior information, and they have a great advantage obvious advantages in the field of image processing. In 2016, Wang pointed out that deep learning is expected to promote the further development of CT imaging technology in the review[2]. With the deepening of the combination of CT imaging and deep learning, deep neural networks (DNN) have been applied in image reconstruction such as low-dose reconstruction and limited-angle reconstruction[3-6]. The above methods greatly improve the quality of the reconstructed image. Deep learning-based methods provide some new ideas to solve the limited-angle problem.

However, the sinogram information has not received great attention in the limited-angle problem. In the sinogram, the intuitive representation of the limited-angle problem is the absence of a portion of the continuous projection data. The lack of projection data is similar to the image inpainting problem which is based on the information in the image to restore the missing parts of the image. In this problem, it is difficult to accurately restore images by traditional methods such as TV-based methods when there is much miss information. Since 2017, the generative adversarial networks (GAN) has achieved good results in image inpainting[7-9]. The generator in GANs is used to learn the probability distribution of the training samples and make the restored images conform to the learned distribution law. Then the discriminator in GANs cannot distinguish whether the image is generated or true.

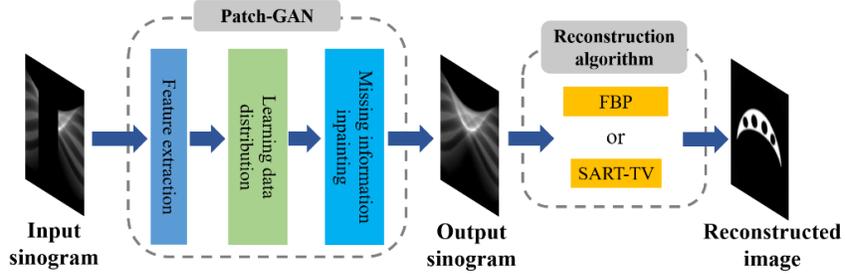

**Fig. 1** The overall workflow of the proposed method.

Inspired by the application of GANs in image inpainting, this paper aims at proposing a sinogram inpainting method for limited-angle CT. The overall workflow of the proposed method is shown in Fig. 1. The sinogram data distribution is learned by training the proposed patch-GAN which can make the sinogram inpainting more accurate. Then, we use the classic CT reconstruction method to obtain the reconstruction image from the generated sinogram. We have preliminary verified the potential of learning the sinogram data distribution by the proposed GAN-based method in very limited scanning angle, such as 60°.

## 2   Method

Recently, Isola et al.[7] proposed an image-to-image network named patch-GAN, which has gained extensive attention. This network has been successfully used in image inpainting, semantic segmentation, and image super-resolution.

In the sinogram inpainting problem, we improved on the architecture of patch-GAN, which can learn a mapping $G: x \to y$ from limited-angle projection data $x$ to output 180° projection data $y'$. The training label of the generator $G$ is 180° full-angle projection data $y$. The generator $G$ is trained to produce outputs that cannot be distinguished from real 180° projection data by an adversarially trained discriminator $D$, which is trained to do as well as possible at detecting the generator's fake projection data. The discriminator $D$ input is a pair of matching data $\{x, y\}$ or $\{x, y'\}$. In the training process, the objective of a patch-GAN can be expressed as

$$L_{patch-GAN}(G,D) = \mathbb{E}_{x,y}\left[\log D(x,y)\right] + \mathbb{E}_{x}\left[\log\left(1 - D(x,G(x))\right)\right], \quad (1)$$

where $x$ is the limited-angle projection data, $y$ is the training label which is the 180° full-angle projection data. For sinogram inpainting tasks, the input and output of $G$ actually share some projection information of limited-angle scans. Therefore, mean absolute error (MAE) between limited-angle projection data and 180° full-angle projection data is introduced into loss function. Then, the total loss function is:

$$L(G,D) = L_{patch-GAN}(G,D) + \lambda \mathbb{E}_{x,y}\left[\|y - G(x)\|_1\right], \quad (2)$$

where $\lambda$ is the weight parameter which is used to balance the two loss functions. During the network training, $G$ tries to minimize the objective function against an adversarial $D$ that tries to maximize the objective function, i.e. $G^* = \arg\min_{G} \max_{D} L(G,D)$.

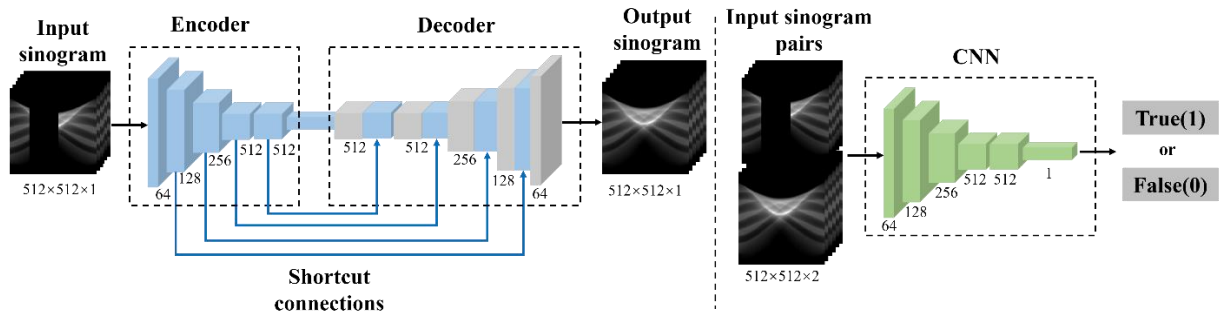

**Fig. 2** Network structure for the generator (left) and discriminator (right).

As shown in Fig. 2 (left), the generator is a fully convolutional network (FCN) which includes encoder and decoder. The encoder extracts sinogram features from the input data by five convolutional layers. The input sinogram image has a uniform size of $512 \times 512$. The first, second and third convolutional layer has 64, 128 and 256 channels of filter kernel size $4 \times 4$, respectively, and the overlapping stride is 2. The fourth and fifth convolutional layer has 512 channels of filter kernel size $4 \times 4$, and the overlapping stride is 2. The activation functions of the five convolution layers are Leaky ReLU with slope 0.2. The purpose of the decoder module is composing the completed sinogram image from the acquired sinogram feature information. The decoder consists of five deconvolutional layers. The first, second, third and fourth deconvolutional layer has 512, 512, 256 and 128 channels of filter kernel size $4 \times 4$, respectively, and the overlapping stride is 2. The final deconvolutional layer has 64 channels of filter kernel size $4 \times 4$, and the overlapping stride is 2. The shortcut connection usually connects the corresponding layer of encoder and layer of the decoder to help the decoder better complete the details of the sinogram. The output image size of the generator is the same as the input $512 \times 512$.

As shown in Fig. 2 (right), the structure of the discriminator is a CNN with inputs as pairs of limited-angle sinogram image and full-angle sinogram image (generated or original). There are six layers in the discriminator. The first three layers are $4 \times 4$ convolutional-ReLU layers with stride 2 and 64, 128, 256 filters, respectively. The fourth and fifth layers are $4 \times 4$ convolutional-ReLU layers with stride 1 and 512 filters. The final layer is a $4 \times 4$ convolutional layer with stride 1 and 1 filter. The output of the discriminator is true or false of matching sinogram pairs, which is equivalent to performing the 0-1 classification.

In this paper, the proposed limited-angle image reconstruction method is divided into the following three steps: First, for training the network, we need to establish a dataset which includes matched pairs of limited-angle sinogram images and 180° full-angle sinogram images. Second, we can input the sinogram of limited-angle scanning into the trained network and obtain the generated 180° sinogram. Finally, because the generated 180° sinogram is complete, we can obtain the reconstruction image by classic methods, such as FBP algorithm and simultaneous algebraic reconstruction technique with total variation regularization (SART-TV)[10] with fewer iterations.

## 3   Experimental Results

The experimental data are acquired from a real clinical dataset which contains 200 cranial cavity $512 \times 512$ images from 5 patients. In order to train the network, the images from 4 patients were selected to generate training samples. For each CT images, we generate 20 limited-angle sinogram images with a 60° angle scan and their corresponding 180° full-angle sinogram images. Poisson noise corresponding to $1 \times 10^5$ photon counts is added to limited-angle sinogram images. We can change the missing angle direction to generate these projection samples. We obtained 3200 pairs of training sinogram images. In our database, each sinogram image has a size of $512 \times 512$. Besides, in order to obtain more samples, each sample is divided into 64 image blocks which have a size of $64 \times 64$. The total number of training samples is 204800, and the training process includes 800 epochs. After training, we test our trained network model on the 60° limited-angle sinogram images which are generated from cranial cavity images of the patient outside the training dataset. In this paper, peak signal to noise ratio (PSNR), root mean squared error (RMSE) and structural similarity index (SSIM) are used to evaluate the reconstruction image.

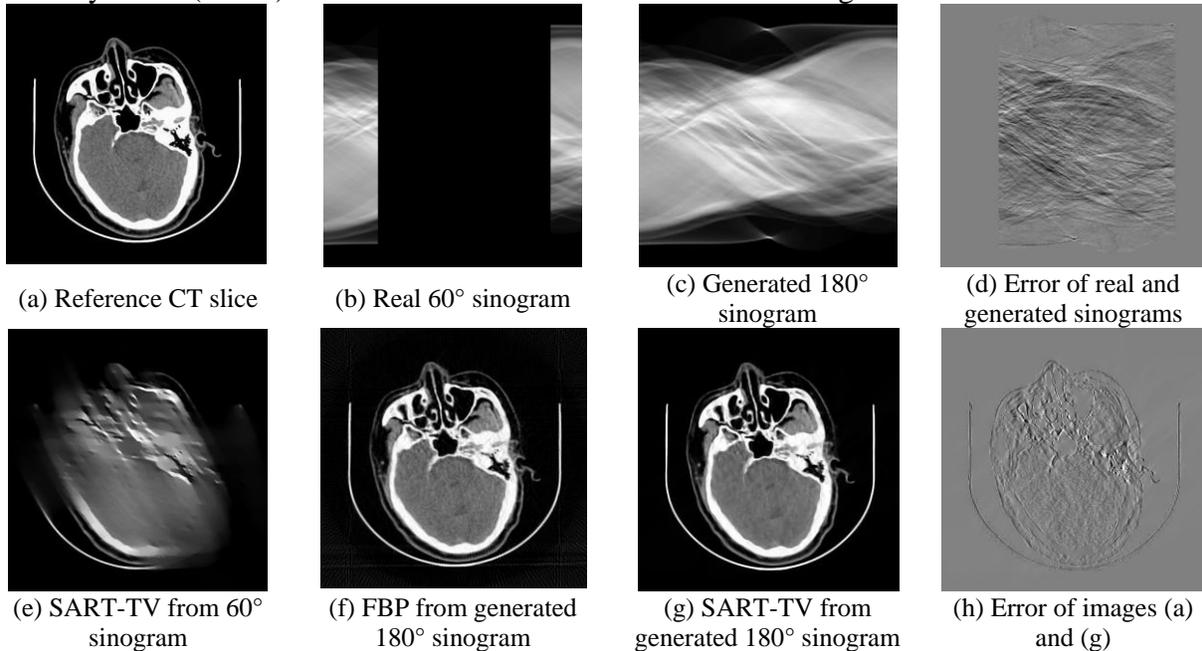

(a) Reference CT slice  (b) Real 60° sinogram  (c) Generated 180° sinogram  (d) Error of real and generated sinograms

(e) SART-TV from 60° sinogram  (f) FBP from generated 180° sinogram  (g) SART-TV from generated 180° sinogram  (h) Error of images (a) and (g)

**Fig. 3** Results of inpainting sinogram and reconstructed images from 60° limited-angle scanning. Display windows of (a)-(c), (e)-(g), (d) and (h) are [0, 1], [0, 0.395], [-0.06, 0.06] and [-0.1, 0.1], respectively.

**Table 1** Evaluation of results reconstructed with different algorithms from generated sinogram data.

|  |  | PSNR | RMSE | SSIM |
|---|---|---|---|---|
| **Generated 180° sinogram** | **FBP** | 25.3285 | 0.0221 | 0.7191 |
|  | **SART-TV** | 30.7205 | 0.0147 | 0.9479 |

Fig. 3 shows the results of sinogram inpainting and reconstruction by the proposed method under 60° limited-angle scanning. We can find that the generated 180° sinogram data is almost visually the same as the real 180° sinogram data and there are fewer errors in the missing scanning angles in Fig. 3 (d). In the reconstruction image from the generated 180° full-angle sinogram data, the serious artifacts caused by missing continuous projection data under limited angle scanning are effectively reduced. This means the proposed method can preliminarily solve the limited-angle problems with smaller angles, such as 60°. However, in the reconstruction image by FBP, we find some noise-like artifacts which may be caused by the subtle errors in sinogram inpainting. Besides, we find these noise-like artifacts can be effectively smoothed after 20 iterations using SART-TV. But there are still some large errors in individual pixels at the edge of the reconstructed image in Fig. 3 (h). Table 1 lists computational results of FBP and SART-TV from generated sinogram data. By comparing PSNR, RMSE, and SSIM, the proposed method realizes better reconstruction results. Therefore, one potential approach to solve this problem is to use some TV-like regularizations. Nevertheless, the results of sinogram inpainting based image reconstruction using the proposed method are promising and can be utilized in solving smaller limited-angle problems.

## 4 Conclusion

In this paper, we proposed a GAN-based sinogram inpainting method which can generate 180° full-angle projection data when the missing angle range is very large. The reconstructed images can be obtained using the traditional FBP algorithm and the classic SART-TV algorithm with dozens of iterations from the projection data generated under 60° limited angle scanning. Experimental results show that serious artifacts caused by 60° limited scanning angle can be suppressed effectively by the proposed method. In the future work, the accuracy of reconstruction results of the actual 60° limited-angle scan has great potential to be improved by designing more efficient reconstruction algorithms for the errors introduced in generated projection data.

*Acknowledgments*

This work is supported by the National Natural Science Foundation of China (NSFC), Nos. 61601518. The contents of this paper are solely the responsibility of the authors and do not necessarily represent the official views of the NSFC. Moreover, as a necessity, the authors declare that all authors have no relevant conflicts of interest to disclose.